\documentclass[final,3p,times,twocolumn]{elsarticle}
\journal{Journal of Systems and Software}
\usepackage{algorithmic}
\usepackage{graphicx}
\usepackage{textcomp}
\usepackage{xcolor}
\usepackage{threeparttable}
\usepackage{tabularx}
\usepackage{colortbl}
\usepackage[inline]{enumitem}
\usepackage{listings}
\usepackage[acronym]{glossaries}
\usepackage{multirow}
\usepackage{siunitx}

\usepackage{xspace}
\usepackage{xcolor}
\usepackage[normalem]{ulem}
\usepackage{ifthen}
\usepackage{amsmath,amssymb}
\usepackage{booktabs}
\usepackage{orcidlink}
\usepackage[capitalise]{cleveref}
\Crefname{figure}{Figure}{Figures}

\definecolor{gray50}{gray}{.5}
\definecolor{gray40}{gray}{.6}
\definecolor{gray30}{gray}{.7}
\definecolor{gray20}{gray}{.8}
\definecolor{gray10}{gray}{.9}
\definecolor{gray05}{gray}{.95}

\newlength\Linewidth
\def\findlength{\setlength\Linewidth\linewidth
	\addtolength\Linewidth{-4\fboxrule}
	\addtolength\Linewidth{-3\fboxsep}
}
\newenvironment{rqbox}{\vspace*{0.2em}\par\begingroup
	\setlength{\fboxsep}{5pt}\findlength
	\setbox0=\vbox\bgroup\noindent
	\hsize=0.95\linewidth
	\begin{minipage}{0.95\linewidth}\normalsize}
	{\end{minipage}
   \egroup
	\textcolor{gray20}{\fboxsep1.5pt\fbox
		{\fboxsep5pt\colorbox{gray05}{\normalcolor\box0}}}
   \endgroup\par\par\noindent
   \normalcolor\ignorespacesafterend}

\RequirePackage[skins,xparse]{tcolorbox}
\NewTColorBox{custombox}{o}{
    colback=gray!10,
	colframe=gray!70,
	left=1.5mm,
	right=1.5mm,
	top=0.1mm,
	bottom=0.1mm,
	fonttitle=\bfseries,
	arc=0mm,
	leftrule=1mm,
	rightrule=0mm,
	toprule=0mm,
	bottomrule=0mm,
	notitle,
	before=\par\medskip\noindent,
    IfValueTF={\small#1}{before upper={\small\textbf{#1: } }}{},
}

\newcommand{\rqa}{$RQ_1$}
\newcommand{\rqb}{$RQ_2$}
\newcommand{\rqc}{$RQ_3$}

\newcommand{\llama}{{Llama}}
\newcommand{\codellama}{{CodeLlama}}

\newcommand{\codegranite}{{Granite Code}}
\newcommand{\rqI}{To what extent can pretrained \glspl{llm} generate \gls{qa} pairs directly from source code for the purpose of code comprehension fine-tuning, and what types of comprehension-relevant questions do they produce?}
\newcommand{\rqII}{How does \gls{kant} compare to a fine-tuned \gls{rag} system (\gls{sota}) in answering repository-specific comprehension questions, with respect to answer quality and inference efficiency?}
\newcommand{\rqIII}{To what extent does \gls{kant} generalize across different underlying \glspl{llm}, and how does its performance vary?}

\graphicspath{{figures/}}

\newcommand{\commented}[1]{}

\newcommand{\eg}{\emph{e.g.,}\xspace}
\newcommand{\ie}{\emph{i.e.,}\xspace}

\usepackage{url}            
\makeatletter
\def\url@leostyle{%
  \@ifundefined{selectfont}{\def\UrlFont{\textsf}}{\def\UrlFont{\small\sffamily}}}
\makeatother
\usepackage{color}
\usepackage{listings}
\usepackage{etoolbox}
\definecolor{stComment}{rgb}{0.5,0.5,0.5}
\definecolor{stString}{rgb}{0.58,0,0.82}
\definecolor{stKeywords}{rgb}{0.21,0.55,0.7}
\definecolor{stNumbers}{rgb}{.5,0,0}

\newtoggle{InString}{}
\togglefalse{InString}
\usepackage{color}
\usepackage{textcomp}
\usepackage{listings}
 
\definecolor{source}{gray}{0.85}
\newcommand{\myCommentStyle}[1]{{\sffamily\color{gray!100!white} #1}}
 
\newcommand{\myStringStyle}[1]{{\sffamily\color{violet!100!black} #1}}
 
\newcommand{\mySymbolStyle}[1]{{\sffamily\color{violet!100!black} #1}}
 
\newcommand{\myKeywordStyle}[1]{{\sffamily\color{green!70!black} #1}}
 
\newcommand{\myGlobalStyle}[1]{{\sffamily\color{blue!100!black} #1}}
 
\makeatletter
\newcommand*\idstyle[1]{%
\expandafter\id@style\the\lst@token{#1}\relax%
}
\def\id@style#1#2\relax{%
\ifnum\pdfstrcmp{#1}{\#}=0%
\mySymbolStyle{\the\lst@token}%
\else%
\edef\tempa{\uccode`#1}%
\edef\tempb{`#1}%
\ifnum\tempa=\tempb%
\myGlobalStyle{\the\lst@token}%
\else%
\the\lst@token%
\fi%
\fi%
}
\makeatother

\lstnewenvironment{code}{%
 \lstset{%
 frame=single,
 framerule=0pt,
 mathescape=false
 }%
 \noindent%
 \minipage{\linewidth}%
}{%
 \endminipage%
}%
\lstnewenvironment{codeWithLineNumbers}{%
 \lstset{%
 frame=single,
 framerule=0pt,
 mathescape=false,
 numbers=left
 }%
 \noindent%
 \minipage{\linewidth}%
}{%
 \endminipage%
}%

\RequirePackage{siunitx}
\newcolumntype{L}{>{\raggedleft\arraybackslash}X}
\newcolumntype{R}{>{\raggedright\arraybackslash}X}

\newlist{compactitem}{itemize}{1}
\setlist[compactitem,1]{label=--, left=0pt, itemsep=0pt}

\definecolor{systemcolor}{gray}{0.3}      
\definecolor{promptcolor}{gray}{0.5}      
\definecolor{placeholdercolor}{gray}{0.9}

\lstdefinelanguage{PromptTemplate}{
  morekeywords={SP, UP},
  sensitive=true,
  morecomment=[l]{//},
  morestring=[b]",
  keywordstyle=\color{systemcolor}\bfseries,      
  stringstyle=\color{promptcolor},   
  commentstyle=\color{gray},
  basicstyle=\color{gray},
  literate={\{}{{{\color{placeholdercolor}\{}}}1  
           {\}}{{{\color{placeholdercolor}\}}}}1  
}

\newacronym{oss}{OSS}{Open-source software}
\newacronym{qa}{QA}{question-answer}
\newacronym{se}{SE}{software engineering}
\newacronym{so}{SO}{Stack Overflow}

\newacronym{llm}{LLM}{Large Language Model}
\newacronym{ft}{FT}{Fine-tuning}
\newacronym{sft}{SFT}{Supervised fine-tuning}
\newacronym{fim}{FiM}{Fill-in-the-Middle}
\newacronym{nlp}{NLP}{Natural Language Processing}
\newacronym{BRNN}{BRNN}{Bidirectional-Recurrent-Neural-Network}
\newacronym{NN}{NN}{Neural-Network}
\newacronym{RNN}{RNN}{Recurrent-Neural-Network}
\newacronym{CNN}{CNN}{Convolutional-Neural-Network}
\newacronym{GRU}{GRU}{Gate-Recurrent-Unit}
\newacronym{LSTM}{LSTM}{Long-Short-Term-Memory}
\newacronym{rag}{RAG}{Retrieval Augmented Generation}

\newacronym{SOP}{SOP}{State-of-Practice}
\newacronym{sota}{SOTA}{State-of-the-Art}

\newacronym{kant}{KANT}{Key Augmented Neural Trigger}

\begin{document}

\begin{frontmatter}
\title{Key-Augmented Neural Triggers for Knowledge Sharing}

\author[1]{Alex Wolf\orcidlink{0000-0002-0964-2224}}
\ead{wolf@ifi.uzh.ch}

\author[1]{Marco Edoardo Palma\orcidlink{0000-0003-3300-4828}}
\ead{marcoepalma@ifi.uzh.ch}

\author[1]{Pooja Rani\orcidlink{0000-0001-5127-4042}}
\ead{rani@ifi.uzh.ch}

\author[1]{Harald C. Gall\orcidlink{0000-0002-3874-5628}}
\ead{gall@ifi.uzh.ch}

\affiliation[1]{organization={Department of Informatics, 
                              University of Zurich},
                city={Zurich},
                country={Switzerland}}

\begin{abstract}
Repository-level code comprehension and knowledge sharing remain core challenges in software engineering. 
Large language models (LLMs) have shown promise by generating explanations of program structure and logic.
Retrieval-Augmented Generation (RAG), the state-of-the-art (SOTA), improves relevance by injecting context at inference time.
However, these approaches still face limitations: 
First, semantic fragmentation across structural boundaries impairs comprehension, as relevant knowledge is distributed across multiple files within a repository.
Second, retrieval inefficiency and attention saturation degrade performance in RAG pipelines, where long, weakly aligned contexts overwhelm model attention. 
Third, repository specific training data is scarce, often outdated, incomplete or misaligned.
Finally, proprietary LLMs hinder industrial adoption due to privacy and deployment constraints.
To address these issues, we propose Key-Augmented Neural Triggers (KANT), a novel approach that embeds knowledge anchors, symbolic cues linking code regions to semantic roles, into both training and inference. 
Unlike prior methods, KANT enables internal access to repository specific knowledge, reducing fragmentation and grounding inference in localized, semantically structured memory. 
Moreover, we synthesize specialized instruction tuning data directly from code, eliminating reliance on noisy or outdated documentation and comments. 
At inference, knowledge anchors replace verbose context, reducing token overhead and latency while supporting efficient, on premise deployment.
We evaluate KANT via: a qualitative human evaluation of the synthesized dataset's intent coverage and quality across five dimensions; 
compare against SOTA baselines across five qualitative dimensions and inference speed; 
and replication across different LLMs to assess generalizability. 
Results show that the synthetic training data aligned with information-seeking needs: over 90\% of questions and answers were rated relevant and understandable; 77.34\%, 69.53\%, and 64.58\% of answers were considered useful, accurate, and complete, respectively.
KANT achieved over 60\% preference from human annotators and a LocalStack expert over the baselines (e.g., 21\% RAG) and notably the expert preferred KANT in over 79\% of cases.
Also, KANT reduced inference latency by up to 85\% across all models.
Overall, KANT demonstrated its effectiveness across all evaluated areas, implying that it is well-suited for scalable, low-latency, on-premise deployments, providing a strong foundation for repository-level code comprehension.

\end{abstract}

\begin{keyword}
large language models\sep deep learning\sep repository-level\sep code comprehension\sep information-seeking 
\end{keyword}

\end{frontmatter}
\section{Introduction}\label{sec:intro}

Code comprehension remains one of the most fundamental, time-consuming, and cognitively demanding tasks in software engineering~\cite{xia2017measuring,hansen2013makes,zi2025would}. 
It requires developers to acquire knowledge about a software system by navigating software artifacts, reading source code, and consulting related documentation~\cite{xia2017measuring}. 
Studies show that developers spend nearly 58\% of their time on such activities, 
often interrupted by frequent context switches between IDEs, browsers, and documentation platforms, which disrupts cognitive flow and amplify effort~\cite{xia2017measuring}. 
Moreover, a significant portion of developer questions focuses on seeking information to understand code behavior (37\%), usage examples (28\%), and locating definitions or instantiations (18\%)~\cite{stolee2025WhyCodeSearch}. 

For instance, a question like {``Can you explain the purpose of the} \texttt{ExecutionFailedEventDetails} {class and how it's used in the codebase?''}  requires
architectural awareness, dependency tracking, and an understanding of relationships across code elements.  Yet, answering such questions is hard because 
relevant knowledge is fragmented across the project structure,
control flow, and type hierarchies, each offering only a partial view of the codebase~\cite{sadowski2015developers,alwis2008Answering,sillito2006Questions}. 
These challenges are further compounded by outdated, incomplete, and inconsistent resources, including forums, documentation, community chats, and unreliable code comments, 
and by the lack of memory cues that help surface relevant information~\cite{rani2023CommentQuality,gu_empirical_2019,fronchetti_contributing_2023}. 
Overall, these challenges point to a core limitation in comprehension workflows: \textit{semantic fragmentation across structural boundaries and a lack of an internal memory mechanism} that connects questions to the relevant regions of the codebase.

Recent work has explored the use of intelligent, integrated assistance through \glspl{llm} 
in supporting 
code comprehension by interpreting program structure, logic, and dependencies through natural language interaction~\cite{Hou2023LargeLM}. 
They can generate human-readable explanations across diverse programming languages~\cite{su2024context,allamanis2018SurveyOfMLForCodeAndNaturalness,ahmad2021PLBART}, 
helping developers navigate and reason about unfamiliar or complex code regions. 
\gls{rag}, \gls{sota} pipelines extend this capability by incorporating additional context, such as external documentation or code snippets retrieved at inference time, enabling more accurate and grounded answers.

However, despite these advances, fundamental limitations remain. 
These models often hallucinate or misinterpret code behavior~\cite{ji2023SurveyOfHallucination,he2024retrieving}, overlook key information in long contexts due to the ``lost in the middle'' effect~\cite{liu2024LostMiddle}, 
and are sensitive to ambiguous or incomplete 
context~\cite{mallen2023trustLLM,he2024retrieving}. 
Therefore, retrieved content can be noisy, irrelevant, or poorly aligned with the user intent \cite{mallen2023trustLLM,he2024retrieving}.
Specifically, \gls{rag} increases prompt length and token consumption due to context fragmentation and attention saturation in the retrieved context, often degrading model performance ~\cite{levy-etal-2024-task}.
These challenges highlight a second core limitation of such pipelines: \textit{retrieval inefficiency and attention saturation}, where verbose and loosely aligned context overwhelms the model's attention window, fragments reasoning, and degrades answer quality~\cite{levy-etal-2024-task,he2024retrieving,liu2024LostMiddle}.

Fine-tuning \glspl{llm} for specific domains or repositories offers a promising alternative to overcome these limitations. However, it requires representative, high-quality datasets that capture the structural and semantic nuances of real-world codebases, which in practice are scarce. 
Public corpora are often incomplete, unrepresentative of real codebases, outdated, or poorly documented~\cite{rani2023CommentQuality,guizani_long_2021,larios_vargas_selecting_2020,tan_wait_2023}, while industrial repositories are typically proprietary and inaccessible. 
Manually curating up-to-date instruction tuning data at scale is costly and labor-intensive. 
While \gls{rag} pipelines and task-specific fine-tuning help mitigate context limitations, they lack built-in mechanisms for context retention without relying on external dependencies.
This leads to a third core limitation:
\textit{repository specific data scarcity and misalignment}, the limited availability of training corpora aligned with the underlying code semantics specific to, and representative of, individual repositories~\cite{rani2023CommentQuality,guizani_long_2021,tan_wait_2023}. 
This scarcity limits fine-tuning and long-term adaptability of comprehension models.

Collectively, these three limitations hinder the adaptability, precision, and efficiency of comprehension systems.
To address them, we propose \gls{kant}, a novel approach that embeds internal \emph{knowledge anchors}, symbolic cues that capture semantic associations between code regions and their roles in the system. 
Unlike traditional approaches that rely on context-based retrieval or documentation, 
\gls{kant} enables models to retrieve repository-specific knowledge internally, improving contextual grounding and token efficiency. 
Rooted in cognitive psychology~\cite{melton1963ImplicationsMemory, reed2003memory}, \gls{kant} reshapes model representations during training to support more stable and distinctive memory segments aligned with a repository's structure.

\gls{kant} tackles repository-level comprehension by embedding knowledge anchors directly into the model. To mitigate repository-specific data scarcity, \gls{kant} generates a synthetic instruction-tuning dataset directly from source code, enabling scalable, up-to-date training without external documentation, manual annotation, or external supervision. 
The dataset comprises \gls{qa} pairs used in a two-stage fine-tuning pipeline. During \gls{fim} training, code regions are augmented with \textit{knowledge anchors}, allowing the model to associate structural context with anchor keys. Subsequently, instruction tuning reuses the same anchors alongside the \gls{qa} pairs. Together, these stages guide the model to form stable associations between semantically related code regions, addressing semantic fragmentation. At inference time, the model is prompted using only the relevant knowledge anchors, without injecting large in-prompt context, thus enabling compact, memory-efficient inputs. This approach addresses retrieval inefficiency and attention saturation by replacing the construction of verbose prompts and external context retrieval with anchor-based memory access.

The approach is validated through a series of qualitative evaluations: assessing the quality and coverage of the generated instruction dataset, comparing comprehension and inference performance against baselines, and evaluating generalizability across model families.

\noindent In summary, the primary contributions of this paper are the following (associated material is provided in the replication package~\cite{replicationpackage}):
\begin{itemize}
\item A validated automatic pipeline for generating instruction-tuning data points directly from source code, serving as a generalizable foundation for fine-tuning downstream models for repository-level code comprehension and knowledge sharing.

\item The introduction of \gls{kant}, a reusable training approach that distills repository-specific knowledge into internal symbolic anchors mined from code, enabling \glspl{llm} to form resilient, context-rich memory representations, reduce inference-time token consumption, and support fast, deployment-friendly, more accurate, and context-related code comprehension.

\item An in-depth qualitative human evaluation of the synthetic instruction tuning dataset (\gls{qa} pairs) generated directly from source code, demonstrating their semantic fidelity, diversity, and relevance to real-world developer information needs.

\item A comprehensive qualitative benchmark comparing \gls{kant} to fine-tuned \gls{rag}-based baselines, showing consistent gains in answer accuracy, completeness, inference latency, and other dimensions across models.
\end{itemize}

The remainder of the paper is organized as follows: \Cref{sec:approach} describes the \gls{kant} approach and training process. 
\Cref{sec:experiments} outlines our experimental design and threats to validity. 
\Cref{sec:results} presents evaluation results. 
We discuss related work in \Cref{sec:relatedWork} and future work in \Cref{sec:futureWork}. Finally, we conclude in \Cref{sec:conclusion}.

\section{Approach}
\label{sec:approach}

This section introduces the \gls{kant} approach for unblocking repository-specific code comprehension in \glspl{llm} through the use of symbolic memory keys, referred to as \emph{knowledge anchors}. 
The \gls{kant} approach is designed to address the three core limitations outlined in \Cref{sec:intro}:
\begin{enumerate*}[label=(\roman*)]

  \item \textit{semantic fragmentation across structural boundaries and a lack of internal memory associations}~\cite{stolee2025WhyCodeSearch,alwis2008Answering,he2024retrieving,de_la_mora_empirical_2018},
  \item \textit{retrieval inefficiency and attention saturation}~\cite{ji2023SurveyOfHallucination,he2024retrieving,mallen2023trustLLM,levy-etal-2024-task,liu2024LostMiddle}, and
  \item \textit{repository-specific data scarcity and misalignment}~\cite{rani2023CommentQuality,guizani_long_2021,tan_wait_2023}.
\end{enumerate*}

Embedding knowledge anchors during training, allows \gls{kant} to equip models with structured internal representations of code regions, 
enabling repository-specific structural reasoning and symbolically grounded memory access, thereby bypassing the need for verbose retrieval, brittle documentation, or prompt-heavy context injection.
This design shifts comprehension from external retrieval to efficient internal memory access, reducing dependence on misaligned auxiliary inputs.

\gls{kant} links each code region with a knowledge anchor that functions as a stable anchor during both training and inference. 
These anchors are embedded into the model alongside source code and \gls{qa} pairs, guiding the model to form localized addressable memory regions. 
Associating each code context with a distinct key would help guide the model to condition its next-token predictions on the relevant code region, increasing the probability that subsequently generated tokens align with the associated region.
This structured separation could help the model distinguish between semantically distinct areas within a repository and retrieve relevant information more effectively by referencing the appropriate key during inference.
The knowledge anchoring enables the model to distinguish between structurally distinct, yet semantically similar regions, thereby reducing fragmentation across file boundaries and promoting location-aware grounding of code semantics.

\gls{kant} builds on recent advances in code-pretrained \glspl{llm}~\cite{roziere2023code} and supports multiple \glspl{llm} families, 
including \codellama~\cite{roziere2023code} and \codegranite~\cite{mishra2024granite}. 
It comprises of three components:
\begin{enumerate*}[label=\alph*)]
  \item synthetic training data generation, 
  \item a knowledge anchoring mechanism, 
  \item and a two-stage key-augmented training process.
\end{enumerate*}

The remainder of this section outlines the components along with the training and inference protocols.

\subsection{Synthetic training data generation}
To address the scarcity of aligned, repository-specific training data, \gls{kant} synthesizes \gls{qa} pairs directly from the relevant source code. 
Unlike existing data generation approaches, such as CS1QA~\cite{lee2022cs1qa}, and ProCQA~\cite{li2024procqa}, 
which derive questions from code comments or community platforms (e.g., GitHub, Stack Overflow), 
\gls{kant} reduces reliance on external sources that are frequently outdated, missing, or poorly aligned with code semantics~\cite{rani2023CommentQuality,guizani_long_2021,tan_wait_2023}. 
This enables the generation of semantically grounded and repository- and task-specific \gls{qa} pairs, with questions such as ``Where is the retry logic implemented?'' or ``What does this function return?'', tailored to comprehension needs~\cite{sadowski2015developers,stolee2025WhyCodeSearch}, and suitable for reuse in downstream tasks like summarization and search, where linking developer queries to specific code regions is essential.

The process begins by partitioning the codebase into token constrained segments, filtering out non-code artifacts (e.g., binaries, media). 
This improves over standard fine-tuning pipelines by tightly coupling training data with the actual code structure, 
directly addressing limitation iii regarding dataset alignment and scarcity.

\begin{figure*}[htb]
  \begin{center}
    \includegraphics[width=0.85\textwidth]{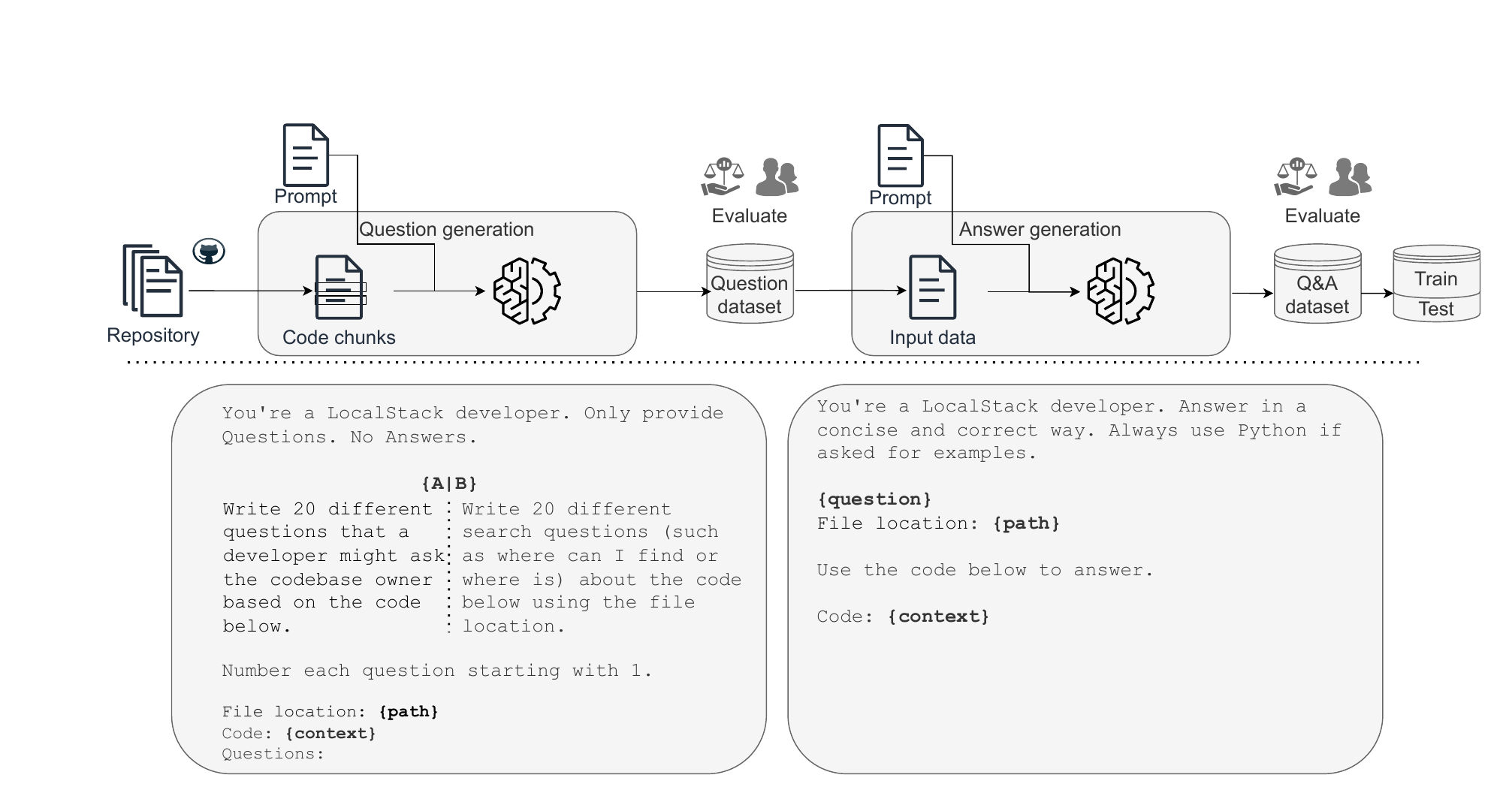}
  \end{center}
  \caption{Training data generation and evaluation pipeline (for \rqa)}\label{fig:promptPipeline}
\end{figure*}

Each code chunk $c_i$ is processed using structured prompts to generate questions $q_i$ and corresponding answers $a_i$ from the same context, as shown in \cref{fig:promptPipeline}. 
This guarantees high alignment between training data and the code itself. 
We formalize this as a generation function $\mathcal{G}(c_i, p_i) \rightarrow (q_i, a_i)$, where $p_i$ denotes the chunk's position within the repository and contributes positional context to the generation process, resulting in a dataset of tuples $(c_i, q_i, a_i, p_i)$, outlined in \cref{tab:dataset_stats}, that encode both semantic and structural cues for training. 
This pipeline reflects practical developer inquiries (e.g., ``Where is the retry logic implemented?'' or ``What does this function return?'')~\cite{sadowski2015developers,stolee2025WhyCodeSearch}, and eliminates the brittleness of comment-based corpora by grounding question generation directly in code semantics.
\begin{table}[htb]
  \centering
  \caption{Dataset statistics}
  \label{tab:dataset_stats}
  \resizebox{\linewidth}{!}{
    \sisetup{table-format=1.0}
\begin{tabular}{
    l S[table-format=6.0] S[table-format=6.0] S[table-format=6.0]
}

\hiderowcolors
\toprule
  \textbf{Dataset Component} & \textbf{Train}  & \textbf{Test}   \\\hline
\midrule
\showrowcolors
\gls{fim} dataset & 2596 & \text{-} & \\
Generated Q\&A pairs & 164508 & \text{-}  \\
Deduplicated Q\&A dataset & 139611 &  15512\\
\bottomrule
\end{tabular}

  }
\end{table}
\subsection{Key-augmented fine-tuning}
\gls{kant} introduces knowledge anchors $ka \in \mathcal{K}$ into the model as memory keys tied to specific code regions.
These anchors function as cognitively inspired cues~\cite{melton1963ImplicationsMemory, reed2003memory}, conditioning the model's behavior such that predictions become more aligned with the semantics of the anchored code region.
Each training example is expressed as a triple $(ka, x, y)$, where $ka$ is a knowledge anchor, $x$ is the input sequence (e.g., code or question), and $y$ is the prediction target.
During training, the anchor conditions the model's internal state such that the probability distribution over the next tokens shifts toward completions that reflect the semantic and structural characteristics associated with the knowledge anchor.
This mechanism enables the model to internalize repository specific behaviors and produce responses that remain consistent with localized semantics, even in the absence of full in-context retrieval.

This mechanism mitigates the following:
\begin{enumerate*}[label=limitation (\roman*)]
\item by guiding the model to differentiate and relate overlapping or distributed code regions through knowledge anchor base associations, and
\item by reducing the need for token intensive retrieval and reducing context length at inference time
\end{enumerate*}

\begin{figure*}[htb]
  \begin{center}
    \includegraphics[width=0.85\textwidth]{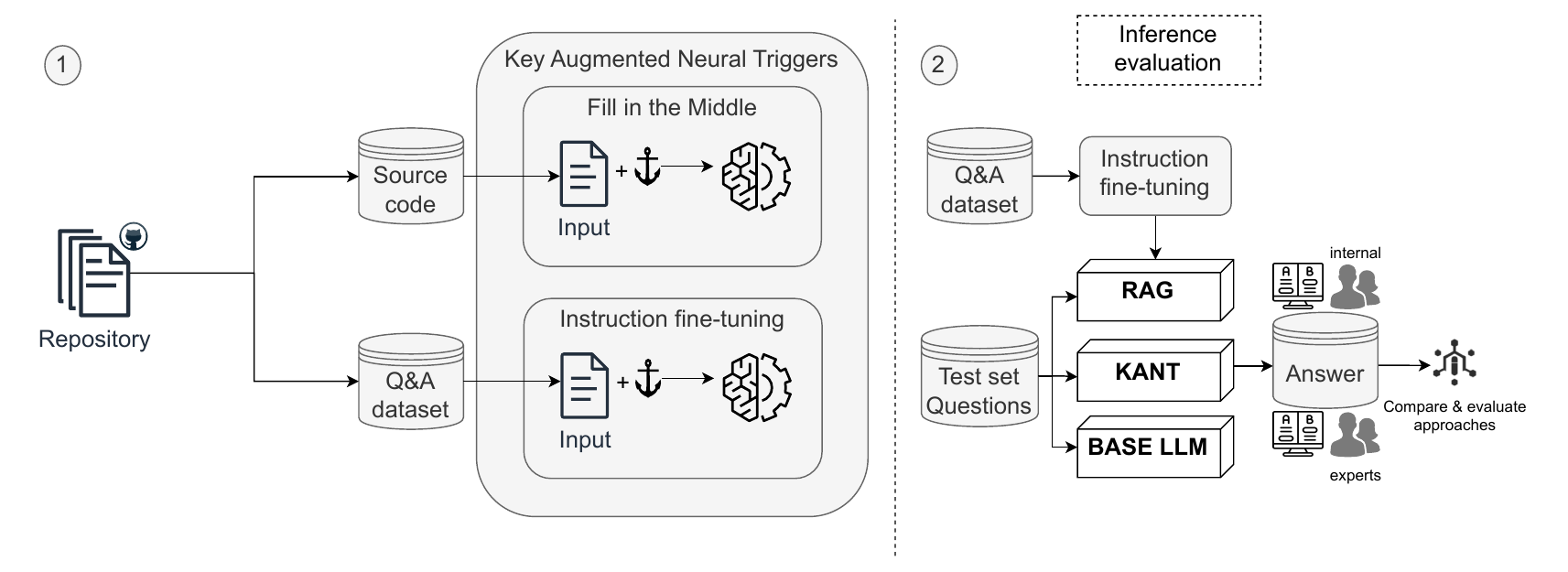}
  \end{center}
  \caption{Two stage fine-tuning and evaluation pipeline (for \rqb~\& \rqc)}\label{fig:trainingPipeline}
\end{figure*}

Training proceeds in two stages, illustrated in \Cref{fig:trainingPipeline}: structural training using \gls{fim} objectives augmented with knowledge anchors, and instruction tuning using knowledge anchor-aware \gls{qa} data. 

\subsubsection*{Fill-in-the-Middle training with knowledge anchors}
In the first stage, the model is trained using a \gls{fim} objective where each file is segmented into prefix $c_{\text{pre}}$, masked span $c_{\text{mid}}$, and suffix $c_{\text{post}}$. 
A knowledge anchor derived from the file path (e.g., \texttt{src/utils/math.py}) is prepended to each sample:
\[
  x = \texttt{[KEY] }ka\ \texttt{[CTX] }c_{\text{pre}}\ \texttt{\textless FILL\textgreater } c_{\text{post}}, 
  \qquad
  y = c_{\text{mid}}.
\]

Anchors are designed to be:
\begin{enumerate*}[label=(\roman*)]
  \item unique per artifact within the repository, \item hierarchically meaningful, reflecting structural boundaries, and \item stable under minor changes.
\end{enumerate*}
This trains the model to learn contextual distinctions even across syntactically similar code and enforces location-specific knowledge anchoring.

\subsubsection*{Instruction tuning with knowledge anchor aware \gls{qa}}

The second stage perform \gls{qa}-style instruction tuning, where each synthetic question-answer pair ($q_i$, $a_i$) is tagged with the same anchor $ka$ from the \gls{fim} stage:
\[
  x = \texttt{[KEY] }ka\ \texttt{[Q] }q,
  \qquad
  y = \texttt{[A] }a.
\]

This conditioning aligns \gls{qa} behavior with structural memory representations established earlier.
By training the model to answer questions in the presence of knowledge anchors, we reduce hallucinations and promote consistency across questions that refer to similar or overlapping content.
Both stages use QLoRA~\cite{dettmers2024qlora} for efficient low-rank adaptation, allowing the approach to scale to large repositories and models under memory constraints.

\subsection{Inference protocol}

At inference time, a user query $q^*$ is paired with $k$ relevant knowledge anchors $ka^*$ retrieved via semantic similarity using a fine-tuned \gls{rag} mechanism.
The model then generates a response conditioned on this lightweight input:
\[
  \texttt{[KEY] }ka^*\ \texttt{[Q] }q^* \rightarrow \hat{a}
\]

This inference structure allows the model to retrieve previously learned internal knowledge tied to knowledge anchors without relying on in-prompt source code or external documentation.
It effectively reduces context size, and thus reduces token consumption resulting in lower latency and mitigating effects of the ``lost-in-the-middle'' phenomenon (\ie second limitation described in \autoref{sec:intro}).
Moreover, it decreases dependencies on external resources, \eg documentation, and mitigating issues related to ambiguous, incomplete, or misleading context (limitation ii). 

Ultimately, \gls{kant} offers a scalable and token-efficient alternative for code comprehension in large and evolving repositories, grounded in knowledge anchoring and self-generated training data.

\section{Experiments}\label{sec:experiments}
This section presents the experimental design used to evaluate the proposed \gls{kant} approach.
We address three research questions that assess the quality of synthetically generated training data across multiple dimensions, the inference performance of \gls{kant} compared to \gls{sota} baselines using qualitative criteria and inference latency, and the generalization across model families.

\begin{rqbox}
    \begin{description}
      \item[\textbf{\rqa}] \rqI
    \end{description}
\end{rqbox}

\noindent To train \glspl{llm} for software engineering \gls{qa} tasks, prior datasets such as CodeQA~\cite{liu2021CodeQA}, CS1QA~\cite{lee2022cs1qa}, and ProCQA~\cite{li2024procqa} derive questions from external sources such as code comments or community forums (e.g., StackOverflow). 
These sources are often outdated, incomplete, or misaligned with code semantics, and require manual annotation or human supervision, which limits scalability and adoption~\cite{rani2023CommentQuality}. 
In this research question, we investigate whether instruction-tuned \glspl{llm} can instead synthetically generate up-to-date and repository specific \gls{qa} pairs directly from source code, 
bypassing the need for external sources. 
Demonstrating such a capability would represent a step toward automating repository-level comprehension pipelines by enabling the systematic construction of training data without manual annotation or supervision.

The focus lies on evaluating the syntactic and semantic integrity of these synthetic \gls{qa} pairs, along with their practical utility for downstream comprehension tasks.
To address this, we generated a dataset of \num{164508} \gls{qa} pairs from the LocalStack repository\footnote{LocalStack: \url{https://github.com/localstack/localstack/commit/85b16b2} v3.7} using prompt based \glspl{llm}. 
LocalStack is a popular open-source tool for offline AWS development and testing, maintained by over \num{570} contributors, with $\sim$\num{60000} GitHub stars and 
a codebase of $\sim$\num{3000} mostly Python files.

To evaluate the quality and diversity of the generated data, we conducted a structured human evaluation on the \gls{qa} pairs. 
We manually analyzed a representative sample of \num{384} \gls{qa} pairs 
with 95\% confidence and a 5\% margin of error~\cite{lohr_sampling_2021}.
To ensure question uniqueness and diversity, we applied pairwise clustering (textual similarity threshold $> 0.8$) using the NLTK library\footnote{\url{https://www.nltk.org/}} prior to sampling, resulting in \num{48112} clusters.

Each \gls{qa} pair was randomly and exclusively assigned to one of the three evaluators ($\approx$126 pairs each) to ensure balanced workload, broad coverage, and reliable quality assessment. 
If a pair was unclear or difficult to evaluate, a second evaluator independently reviewed it. In case of disagreement, a third evaluator reviewed and reached the decision using majority voting mechanism.

Each pair was evaluated along two complementary dimensions:
\begin{enumerate*}[label=(\roman*)]
\item \textit{Intent classification:} To analyze the purpose behind each question, we annotated its underlying intent using the taxonomy proposed by \citeauthor{liu2021CodeQA}~\cite{liu2021CodeQA}, which builds on prior work categorizing developer questions by functionality, purpose, and implementation, to support comprehension tasks~\cite{lee2022cs1qa,allamanis2013and,liu2021CodeQA}.
This schema captures a broad range of developer information needs. 
We extended it where needed to reflect question types specific to our dataset, enabling structured analysis of intent diversity in the generated \gls{qa} pairs.

\item \textit{QA quality assessment:} Each \gls{qa} pair was independently rated using binary criteria, three for questions and five for answers, where 1 indicates the criterion was satisfied and 0 otherwise:
relatedness (the question is relevant to the code and the answer addresses the question),
usefulness (the question / answer provides meaningful insights
for comprehension),
understandability (the phrasing of question / answer is clear and unambiguous),
accuracy (the answer is factually correct with respect to the source code), and
completeness (the answer contains all necessary details).
\end{enumerate*}

This evaluation provides empirically show the feasibility and effectiveness of generating synthetic \gls{qa} pairs from source code and validates the resulting dataset for supporting the \gls{kant} approach.
\begin{rqbox}
\begin{description}
\item[\textbf{\rqb}] \rqII
\end{description}
\end{rqbox}

This question investigates whether \gls{kant}, by embedding knowledge anchors, improves a model's ability to answer repository-level comprehension questions. 
LLMs, when used for software repositories, often struggle to reason over fragmented information (semantic fragmentation across structural boundaries)~\cite{liu2024LostMiddle,mallen2023trustLLM,ji2023SurveyOfHallucination,he2024retrieving}. While \gls{rag} remains \gls{sota} in many tasks, it relies on token-intensive retrieval methods that introduce latency and reduce scalability.
To this end, we evaluate whether \gls{kant} offers measurable gains in answer quality and inference efficiency compared to (i) a \gls{sota} \gls{rag} pipeline and (ii) a base \gls{llm} without adaptation. 

To answer this, we held out 10\% of the generated dataset as a test set and used the remaining 90\% for training. 
We then used the test set questions to compare the performance of three configurations based on the \codellama-7B~\cite{roziere2023code} model: 
the \gls{kant} model fine-tuned with knowledge anchors as described in \cref{sec:approach}, a \gls{sota} \gls{rag} approach using fine-tuned AoE embeddings~\cite{li-li-2024-aoe} and ChromaDB for top-$k$ semantic retrieval ($k = 5$), 
and the original base model without any adjustments. 
To ensure a fair comparison, we adopted the original hyperparameter settings from prior work~\cite{roziere2023code,li-li-2024-aoe} wherever applicable. However, we reduced the batch size to 16 across all configurations to accommodate GPU memory constraints. All models were trained until convergence or for a maximum of 30 epochs; further details are provided in \cref{tab:hyperparams}.

\begin{table}[htb]
\centering
\caption{Hyperparameter settings used in model comparisons}
\label{tab:hyperparams}
\sisetup{table-format=1.0}
\begin{tabular}{
     l r
}

%

\hiderowcolors
\toprule
  \textbf{Parameter} & \textbf{Values} \\\hline
\midrule
\showrowcolors
  \gls{fim} epochs & 30 \\
  Instruction epochs & 5 \\
  \gls{fim}-rate & 0.5\\
  top-k & 5 \\
  context chunk size & 3000 \\
  context window size & 4000 \\
  batch size & 16 \\

\bottomrule
\end{tabular}

\end{table}
We evaluated the generated answers from each system on the held-out test set using a structured human evaluation protocol.
A statistically representative sample of 100 model outputs was selected based on a power analysis for paired two-tailed t-tests ($\alpha$ = 0.05, effect size = 0.3, power = 0.8)~\cite{lohr_sampling_2021}.
Each generated answer was independently rated by one of three human evaluators, randomly assigned to ensure unbiased distribution and without access to the evaluations of their peers.
In parallel, a domain expert from the LocalStack team performed an independent assessment of the same set of generated answers, also blinded to the evaluators scores to avoid confirmation bias.
Each response was evaluated on a 5-point Likert scale (1 = Low, 5 = High) across six dimensions, similar to the quality assessment used to judge the questions: preference (among \gls{kant}, \gls{rag}, and Base), relatedness, understandability, usefulness, accuracy, and completeness. 
This evaluation design enables a rigorous comparison of model performance by capturing the quality of the generated answers, their consistency across similar questions, and the effectiveness of each system in addressing repository-level comprehension tasks.

\begin{rqbox}
\begin{description}
\item[\textbf{\rqc}] \rqIII 
\end{description}
\end{rqbox}

\noindent This question investigates the extent to which the benefits of the \gls{kant} approach generalize across model families of \glspl{llm}, which differ in training data scale, pretraining recency, and hyperparameter configurations. Although, prior work has demonstrated that adapting \glspl{llm} to code-specific tasks can yield performance gains~\cite{ahmad2021PLBART,Hou2023LargeLM,mishra2024granite}, these gains are often sensitive to a model’s pretraining distribution, tokenization scheme, and context window size, raising concerns about robustness and generalizability~\cite{aryabumi2024code,dagan2024Tokenizer,ding2024LongRoPE}. Accordingly, we assess whether the improvements in answer quality and inference efficiency observed in \rqb{} hold when applied to a newer and more capable model. Demonstrating such generalization would provide evidence for the robustness of \gls{kant}, reinforcing the generalizability of the knowledge anchoring approach, and highlighting its potential for adaptation across future model families.

To this end, we replicated the experimental setup from \rqb{}, replacing \codellama{} with \codegranite-8B~\cite{mishra2024granite}, a \gls{llm} trained on a larger and more recent corpus comprising approximately 4.5 trillion tokens of code and technical content. For evaluation comparability, prompt lengths were standardized across models. All other aspects including training hyperparameters for both \gls{kant} and \gls{rag}, inference protocols, and human evaluation procedures were kept fixed to enable a controlled and isolated comparison of outcomes.

\section{Results}\label{sec:results}
This section presents the results obtained from the experiments conducted in \Cref{sec:experiments}. 
For each research question, we analyze our approach's output and evaluate it using appropriate validation methods. 
Our evaluation focuses on three key aspects: (\rqa) synthetic training data quality, (\rqb) the effectiveness and efficiency of the \gls{kant} approach in comparison to \gls{rag}, 
and (\rqc) its generalization across different model families.

\subsection{\rqa~: Quality of synthetic training data}

We evaluated 384 randomly sampled \gls{qa} pairs from our generated data and assessed them along five binary criteria: relatedness, understandability, usefulness, accuracy, and completeness. 
Each pair was independently rated by a human evaluator.

The generated questions demonstrated consistently strong performance across all evaluation dimensions. 
Specifically, we found that 97.66\% of questions were related to the associated code context, 95.57\% were understandable, and 88.02\% were useful for comprehension or development purposes. 
These results indicate that the \gls{llm}-generated questions generally target meaningful aspects of the code and are phrased in a way that aligns with typical developer questions.
For example, the question ``How does the `Gateway` class handle SSL certificates?'' was rated positively across all dimensions. 
It addresses common developer concerns around security configuration, implementation specific details, and operational behavior, and was judged as relevant, understandable, and useful. 

Compared to the questions, the quality of the generated answers showed greater variance. 
We found that 97.66\% of answers were related to their corresponding questions and 90.36\% were understandable, only 77.34\% were considered useful. 
Accuracy and completeness were rated lower, at 69.53\% and 64.58\% respectively. 
These findings suggest that although the answers were generally on topic, they often lacked depth or failed to capture relevant details, particularly when the question required a broader understanding of the codebase.
For example, in response to the previous question about SSL certificate handling, the answer misattributed responsibility, fabricated non-existent API calls, and omitted critical details such as fallback mechanisms and caching behavior.
Consequently, the response was judged incomplete and only partially accurate, illustrating the model's difficulty in synthesizing information that spans multiple interacting components.

To understand the purpose behind the synthetic \gls{qa} pairs,
we annotated each question using the intent taxonomy described in \cref{sec:experiments}. 
As shown in \cref{tab:qaTaxonomy}, most questions focused on code functionality, followed by explanation and purpose.
This distribution closely aligns with findings from prior empirical studies~\cite{stolee2025WhyCodeSearch,sadowski2015developers}, which report that developers most frequently ask questions about what code does, how it works, and why it was written a certain way.
We observed a similar trend in interrogative forms. Most questions began with ``how'', followed by ``what'', while forms such as ``why'', ``where'', and ``who'' appeared only occasionally. This pattern is consistent with the nature of source code, which more readily supports procedural and factual queries than those requiring design rationale or authorship history. Prior studies have also shown that ``how'' questions are among the most common in software development settings~\cite{allamanis2013and,rani2021developers,beyer2020kind,sadowski2015developers,stolee2025WhyCodeSearch}.
The prevalence of functionality-oriented questions, such as ``How is retry logic implemented in this module?'', reflects common developer concerns and confirms that the synthetic data captures information seeking questions observed on platforms such as Stack Overflow~\cite{allamanis2013and,beyer2020kind,rani2021developers}.

\begin{table}[htb]
\centering
\caption{Distribution of question types and intents in the evaluated dataset}
\label{tab:qaTaxonomy}
\sisetup{table-format=1.0}
\begin{tabular}{
     l S[table-format=3.0] | l S[table-format=3.0] 
}

%

\hiderowcolors
\toprule
  \textbf{Category} & \textbf{Count} & \textbf{Type} & \textbf{Count} \\\hline
\midrule
\showrowcolors
functionality &             146 &how	  &    248 \\
explanation &              93  &what	  &    90 \\
purpose &                   91  &explain & 42 \\
property &                  65  &why	  &    8 \\
workflow &                  58  &where	  &    7 \\
example usage &             34  & describe &	4 \\
programming concepts &      18  &which	  &    4 \\
relationship &              13  &when	  &    4 \\
referencing &               10  &whom	  &    1 \\
reasoning &                 6   &who	  &    1 \\
best practices &            4  && \\
knowledge recall &          4 && \\
other &                     11 &&\\
\bottomrule
\end{tabular}

\end{table}

These results confirm that \glspl{llm} can generate high-quality code-specific questions that align well with information seeking questions also observed on community platforms like Stack Overflow~\cite{allamanis2013and,beyer2020kind,rani2021developers}.
Although the answers are often relevant, accurate, and understandable, their usefulness and completeness often suffer due to restricted code visibility. That is, the base model operates over isolated code segments that may omit critical dependencies, related definitions, or higher-level usage contexts necessary to synthesize complete and precise answers. 
For example, answering questions about architectural roles, control flow behavior, or configuration logic may require reasoning across multiple files or modules, which are not visible to the model during generation.
Despite these limitations in answer completeness, the overall quality of the synthetic \gls{qa} pairs demonstrates their feasibility as a source of grounded, code-specific training data. 
These findings underscore the need for improved generation mechanisms that account for broader structural and semantic context, such as knowledge anchors to link questions and answers to specific code regions and improve internal recall of repository-level semantics.
\noindent
\begin{custombox}[\rqa~ -- In summary]
The evaluation demonstrates that the synthetic questions are consistently relevant, well-phrased, and aligned with typical developer concerns observed in prior work~\cite{stolee2025WhyCodeSearch,sadowski2015developers}. 
However, answer quality varies more substantially, with notable deficiencies in completeness and usefulness due to restricted code visibility. 
These findings establish the viability of \gls{llm}-generated \gls{qa} pairs directly from code for fine-tuning, while also underscoring the need for more context-aware methods such as \gls{kant} to improve coverage and depth.
\end{custombox}
\subsection{\rqb: \gls{kant} vs. \gls{rag}: Answer quality and inference efficiency}
To evaluate the impact of knowledge anchors, we compared three configurations based on the \codellama-7B model: a base pretrained model without fine-tuning, a \gls{sota} \gls{rag} model fine-tuned on the same \gls{qa} data, 
and the \gls{kant} model fine-tuned using the procedure described in \cref{sec:approach}. 
All systems were evaluated on the same held-out test set.

Human evaluators consistently preferred \gls{kant} over competing approaches. Across all responses, 62\% of answers generated by \gls{kant} were favored, compared to 21\% for \gls{rag}. The remaining 17\% were either rated in favor of the base model or marked as undecided.
Undecided ratings occurred in two primary scenarios: either all responses were equally unhelpful, such as in response to speculative or poorly posed questions (for example, ``How does the Execution class’s Id property relate to the Id property in the State class?'', where no such relation exists in the code), or all answers were judged equally correct. This second case was rare and typically occurred with simpler questions that could be answered from a single class or file. One such example is, ``What is the name of the table for single metrics in your database?'', where all approaches gave satisfactory answers.
\cref{tab:expert-vs-general-llama} reports the full breakdown, including ratings from a LocalStack domain expert.
The expert preferred \gls{kant} in 87\% of cases, compared to 10\% for \gls{rag}, with the remainder assigned to the base model or marked undecided.

\begin{table}[htb]
\centering
\caption{Answer preferences by approaches: Evaluators vs. Expert}
\label{tab:expert-vs-general-llama}

\sisetup{table-format=2.0}
\begin{tabular}{
     l lrrr
}

\hiderowcolors
\toprule
\multirow{2}{*}{\textbf{Preference}} & \multicolumn{2}{c}{\textbf{CodeLlama}} & \multicolumn{2}{c}{\textbf{\codegranite}} \\
\cmidrule(lr){2-3} \cmidrule(lr){4-5}
                                   & \textbf{Eval.} & \textbf{Exp.}  & \textbf{Eval.} & \textbf{Exp.} \\\hline
\midrule
\showrowcolors
\gls{kant} & 62\% & 87\% & 62\% & 79\% \\
\gls{rag}& 21\% & 10\%  & 17\% & 16\% \\
Base & 5\% & 3\%  & 7\% & 2\% \\
Undecided & 12\% & \text{-} & 14\% & 3\%\\
\bottomrule
\end{tabular}

\end{table}

\begin{figure}[hbt]
\centering
\includegraphics[width=\linewidth]{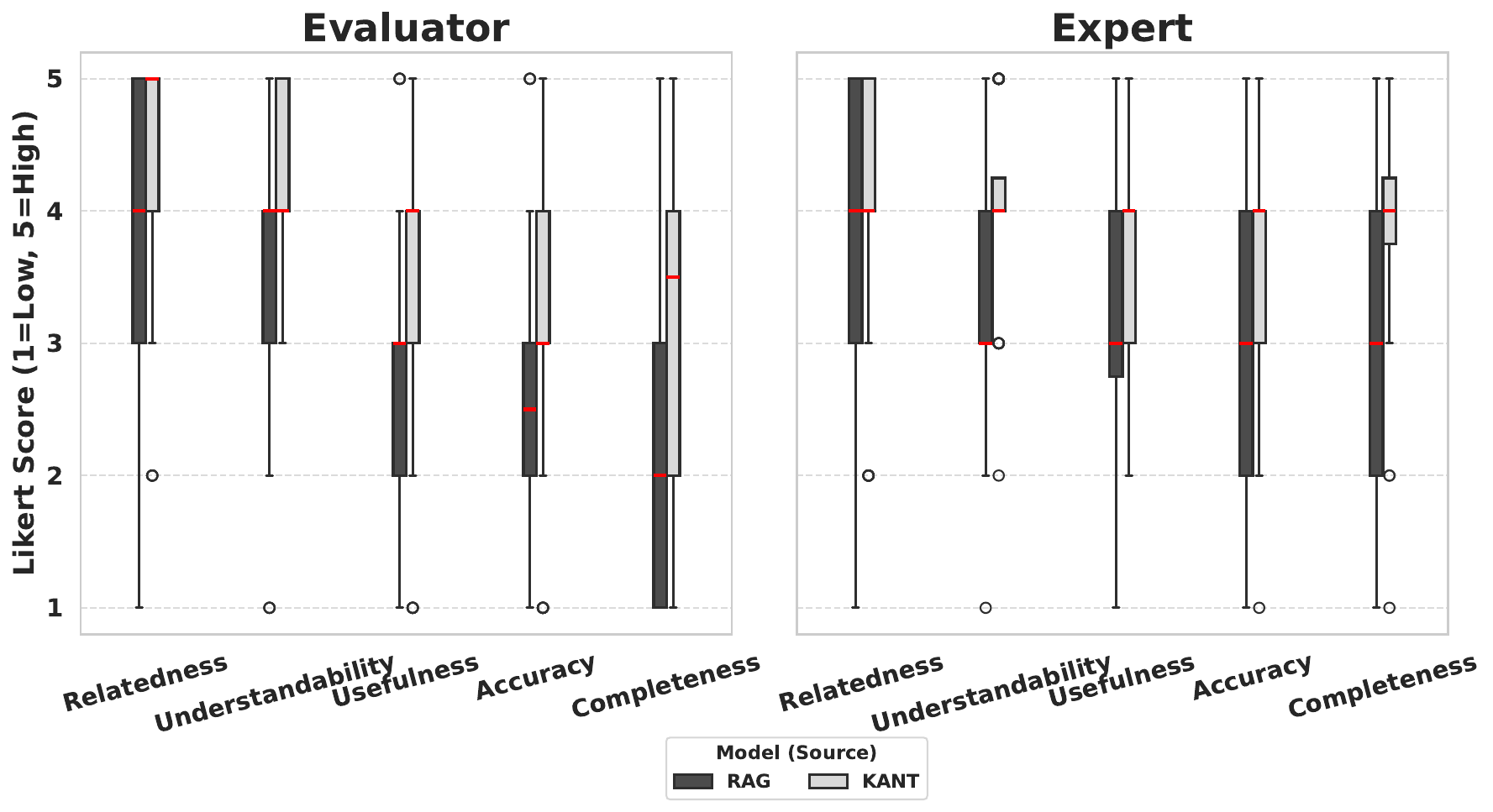}
\caption{Likert scale ratings for \codellama-based models.}
\label{fig:likert-codellama}
\end{figure}

In addition to preference ratings, each answer was scored on a 5-point Likert scale across five evaluation dimensions: relatedness, understandability, usefulness, accuracy, and completeness. 
As shown in \cref{fig:likert-codellama}, \gls{kant} outperformed the \gls{rag} baselines across all criteria.
In particular, knowledge anchoring led to gains in completeness. 
The median completeness rating increased from $2.0 \pm 1.057$ in the \gls{rag} configuration to $3.5 \pm 1.128$ for \gls{kant}. 
Similarly, accuracy improved, with the median rising from $2.5 \pm 1.166$ to $3.0\pm 1.061$, and usefulness from $3.0\pm 1.139$ to $4.0\pm 1.078$.
These results suggest that \gls{kant} helps the model generate more precise and contextually complete responses.
In the Likert scale assessment, the expert rated \gls{kant}'s answers with a median completeness of $4.0\pm 0.803$, accuracy of $4.0\pm 0.838$ and usefulness of $4.0\pm 0.792$, 
compared to \gls{rag}'s scores of $3.0\pm 0.893$, $3.0\pm0.978$, and $3.0 \pm 0.961$, respectively.

\begin{figure}[thb]
\centering
\includegraphics[width=0.8\linewidth,height=0.7\linewidth]{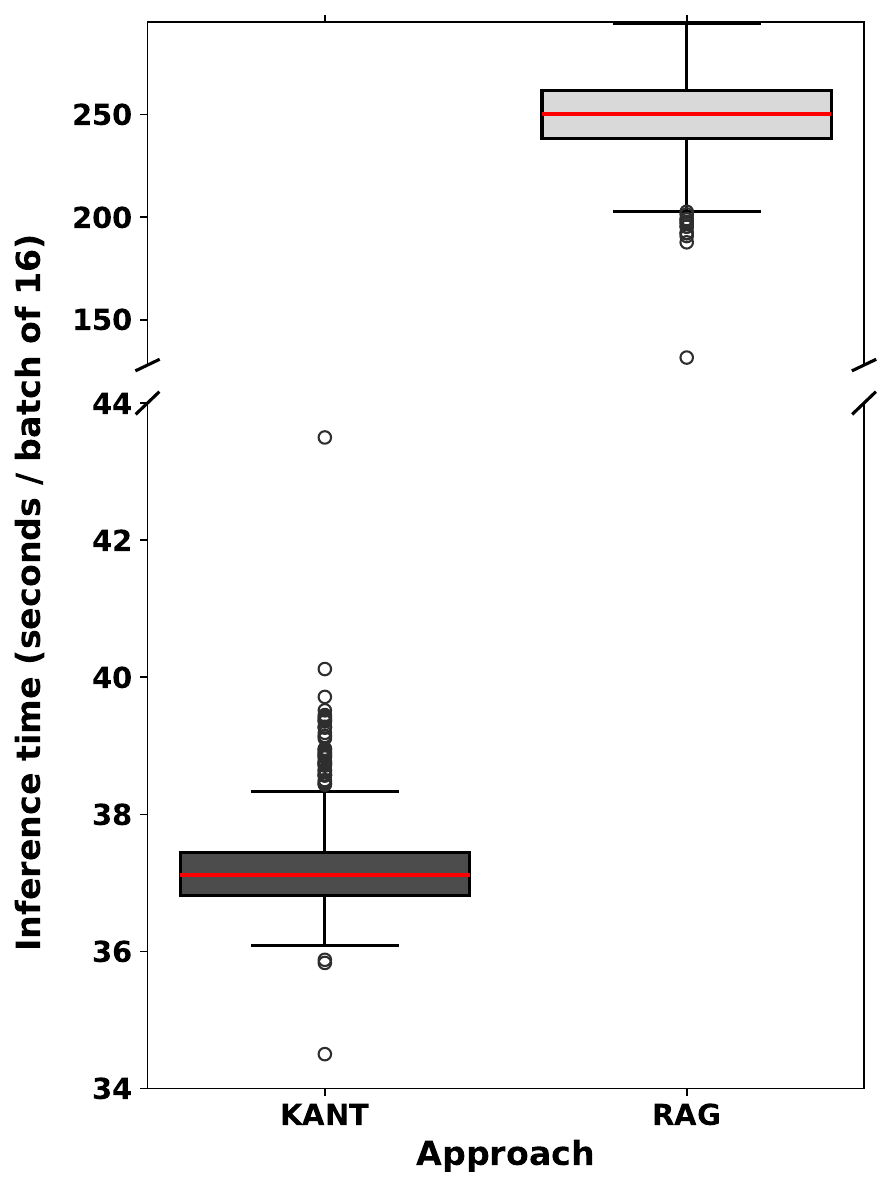}
\caption{Average inference time using \codellama. Lower is better.}
\label{fig:inference-speed-llama}
\end{figure}

Regarding the inference speed, \gls{kant} achieved faster inference, where the average response time per batch of 16 queries was 37.2 seconds, compared to 248.8 seconds for \gls{rag}. 
This represents an approximately 85\% reduction in latency. 
This speedup is primarily due to \gls{kant}'s memory-based conditioning, which avoids the runtime overhead of full-context retrieval, as shown in \cref{fig:inference-speed-llama}.

\begin{custombox}[\rqb~ -- In summary]
\gls{kant} consistently outperforms both the \gls{rag} and base configurations in answer quality, as confirmed by human and expert preference judgments and Likert scale ratings across all dimensions. 
Inference latency was also significantly lower, with an 85\% reduction compared to \gls{rag}. 
These results demonstrate that \gls{kant} is both an effective and efficient approach for repository-level code comprehension.
\end{custombox}
\subsection{\rqc: Generalizability across model families}

To evaluate whether \gls{kant}'s benefits generalize beyond a single model family, we repeated the \rqb~ experiments using \codegranite-8B as the underlying model. 
Both \gls{kant} and \gls{rag} were retrained on the same synthetic \gls{qa} dataset and evaluated on the same test set.

Results remained consistent with those observed for \codellama. 
Human evaluators again preferred \gls{kant} in 62\% of cases, compared to 17\% for \gls{rag}, with the rest favoring the base model or indicating no preference. 
The LocalStack expert showed a similar trend, preferring \gls{kant} in 79\% of cases and \gls{rag} in 16\%.
This suggests that knowledge anchoring remains effective across model variants with varied training setups.

\begin{figure}[thb]
\centering
\includegraphics[width=\linewidth]{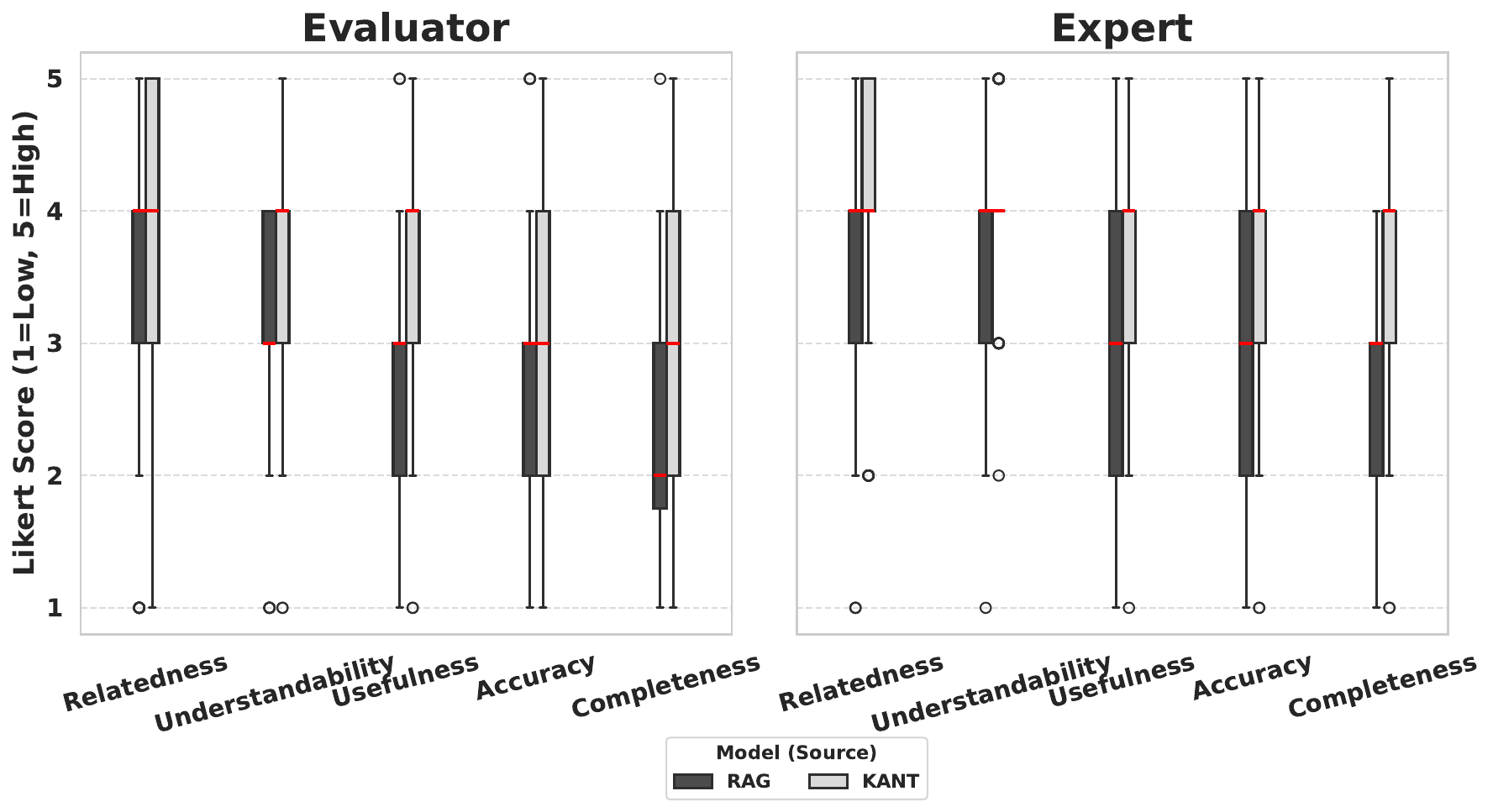}
\caption{Likert scale ratings for \codegranite-based models.}
\label{fig:likert-granite}
\end{figure}

Likert scale ratings further support this finding. 
As shown in \cref{fig:likert-granite}, \gls{kant} achieved higher or comparable scores across all evaluation dimensions. 
Notably, median completeness improved from $2.0\pm0.955$ for \gls{rag} to $3.0\pm1.087$ for \gls{kant}, 
while usefulness increased from $3.0\pm1.029$ to $4.0\pm1.078$. 
Accuracy scores remained stable between models, both with a median of $3.0$, although \gls{kant} showed slightly greater variance ($\pm1.175$ vs. $\pm1.152$), 
indicating a broader range of output specificity.
Similarly, the LocalStack expert assigned \gls{kant} a median completeness score of $4.0\pm 0.905$, accuracy of $4.0\pm0.863$, and usefulness score of $4.0\pm 0.89$, 
compared to \gls{rag}'s scores of $3.0\pm 0.787$, $3.0\pm 0.935$, and $3.0\pm 0.902$, respectively.

The relative speedup of $\approx 80.9\%$ was comparable to the one observed with \codellama, reinforcing the generalizability of its efficiency benefits.

\begin{figure}[thb]
\centering
\includegraphics[width=0.8\linewidth,height=0.7\linewidth]{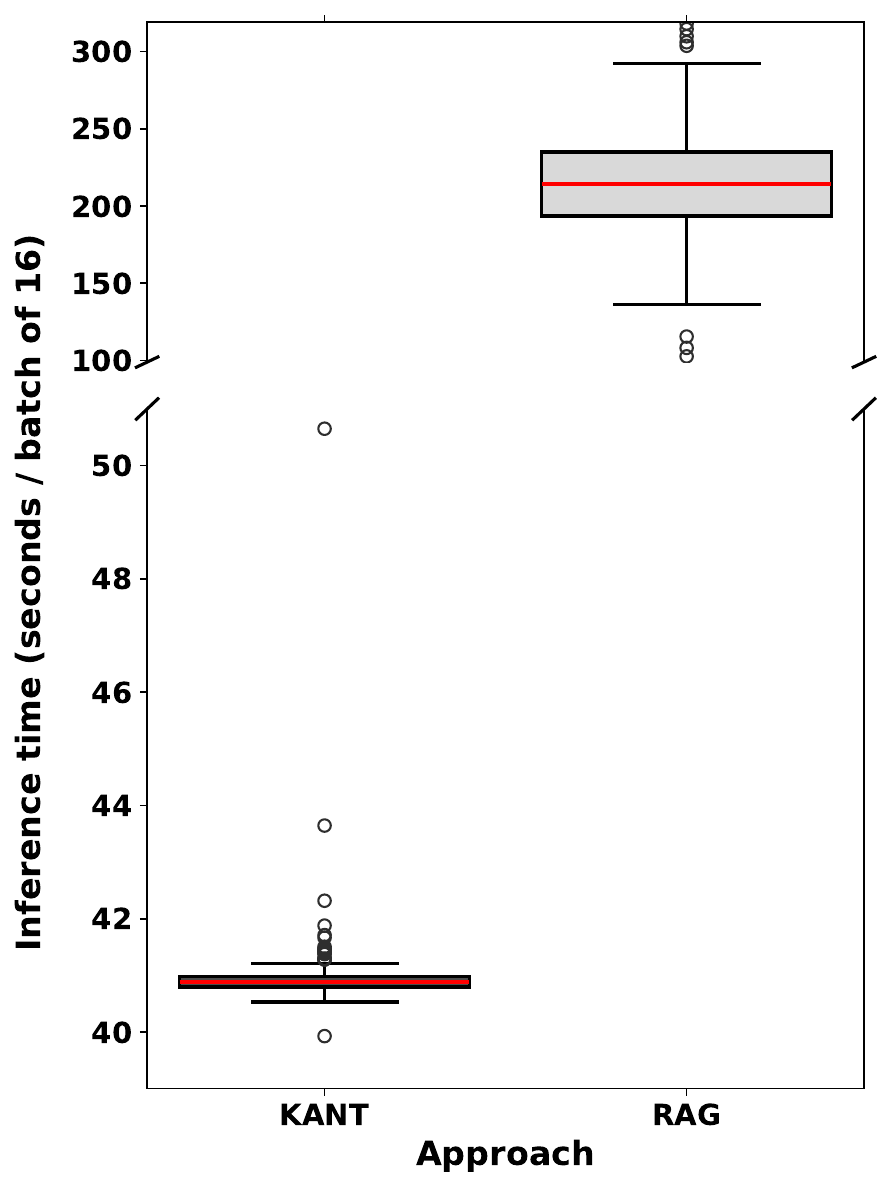}
\caption{Average inference time using \codegranite. Lower is better.}
\label{fig:inference-speed-granite}
\end{figure}

\gls{kant} also maintained its inference time advantage when deployed with \codegranite, as visualized in \cref{fig:inference-speed-granite}. 
The average response time per batch of 16 queries was 80.9\% lower than that of the \gls{rag} configuration. 
This result aligns with the latency reduction observed in the \codellama experiments and reflects the efficiency of knowledge anchoring, 
which removes the need for dynamic context retrieval and prompt expansion.

These results suggest that the design principles underlying \gls{kant}, especially the use of knowledge anchor-based memory for grounding, are robust to changes in model architecture and remain effective with larger or differently structured models.

\noindent\begin{custombox}[\rqc~ -- In summary]
\gls{kant} generalizes effectively across model families. 
When evaluated with \codegranite-8B, it maintained its performance advantages in both answer quality and inference latency. 
Improvements in completeness and usefulness were particularly pronounced, and evaluator preferences mirrored those observed with \codellama. 
These results confirm that knowledge anchoring provides a robust and transferable mechanism for enhancing code comprehension across modern \glspl{llm}.
\end{custombox}

\section{Related work}\label{sec:relatedWork}

\textit{Developer information needs and code search.}
Prior work has extensively examined the types of questions developers ask during software development, particularly in the context of debugging, maintenance, and comprehension tasks~\cite{ko2007information,latoza2010hard,fritz2010using,sillito2006Questions,alwis2008Answering}.
These studies show that developers often struggle to answer complex, context-dependent questions about code, especially those requiring reasoning across files or components.
For instance, \citeauthor{ko2007information}~\cite{ko2007information} highlight the significant effort developers invest in understanding unfamiliar code, while \citeauthor{sillito2006Questions}~\cite{sillito2006Questions} catalog a diverse questions, many involving relationships among distributed elements.
\citeauthor{alwis2008Answering}~\cite{alwis2008Answering} describe such inquiries as ``conceptual queries'' that are inherently difficult to resolve.
Recent findings confirm that even with access to \gls{llm}-based tools, developers continue to ask explanation-seeking questions about code behavior, intent, and design that current models fail to answer effectively~\cite{haque2024information,richards2024you,stolee2025WhyCodeSearch}.
Although sometimes treated as a standalone task, code search is increasingly seen as central to code comprehension, as shown by studies that examine how developers use search to acquire knowledge about software systems~\cite{allamanis2018SurveyOfMLForCodeAndNaturalness,xia2017measuring,stolee2025WhyCodeSearch}.
It supports activities such as code reuse, bug localization, and exploration~\cite{sadowski2015developers,di2023code}, and must operate over language-specific syntax, semantic variation, and abstract program representations~\cite{husain2019codesearchnet}.
Field studies further show that developers continue to struggle with identifying code usage, architectural relationships, and rationale~\cite{stolee2025WhyCodeSearch}.

\gls{kant} addresses these gaps by jointly training on comprehension and search-style questions anchored to specific code regions, enabling models to reason over repositories more effectively and capture relationships  spanning multiple components and files.

\textit{Question answering datasets for comprehension.}
To support automated code comprehension and \gls{qa} tasks, benchmark datasets have emerged to evaluate both retrieval and question answering capabilities, often categorizing questions by intent, (\eg functionality or purpose~\cite{allamanis2013and,liu2021CodeQA,lee2022cs1qa}).
CodeSearchNet~\cite{husain2019codesearchnet}, focuses on text-to-code retrieval, while datasets like CodeQA~\cite{liu2021CodeQA}, CS1QA~\cite{lee2022cs1qa}, ProCQA~\cite{li2024procqa}, and CoReQA~\cite{chen2025coreqa} derive questions from code comments, GitHub issues, or community platforms like Stack Overflow.
While these datasets are valuable for training and evaluation, they suffer from two key limitations.
First, they rely heavily on external artifacts such as comments, documentation, GitHub issues, or forum posts, which are often outdated, incomplete, or misaligned with the actual source code~\cite{rani2023CommentQuality,guizani_long_2021,larios_vargas_selecting_2020,tan_wait_2023,gu_empirical_2019}.
Second, the resulting \gls{qa} pairs tend to be highly localized, typically scoped to a single function or small code segment~\cite{chen2025coreqa} and thus fail to capture broader semantic and structural reasoning across files or modules, which is essential for addressing the diverse set of developer information needs.

The \gls{kant} approach fundamentally differs from prior works. 
Instead of mining documentation or external sources, \gls{kant} generates synthetic \gls{qa} pairs directly from the code itself using structured prompt templates for comprehension and search. 
This enables up-to-date, internally consistent training data that reflects divers developer information needs. 
Furthermore, \gls{kant} uses these \gls{qa} pairs not just for evaluation, but to fine-tune \glspl{llm} using embedded knowledge anchors that link questions to specific code elements. 

\textit{Use of \glspl{llm} in code comprehension}
\glspl{llm} have been applied to software tasks such as code explanation~\cite{richards2024you}, program repair~\cite{nashid2023retrieval}, and interactive development~\cite{nam2024using}. 
Several studies focus on their use in QA settings, particularly using \gls{rag} to combine retrieved information with generative models~\cite{haque2024information,hicke2023chata}. 
However, \gls{rag} performance depend on retrieval quality, which may degrade in the absence of comprehensive documentation or when applied to proprietary code. 
Moreover, they often inflate context windows with long, noisy inputs, amplifying the \emph{lost-in-the-middle} effect~\cite{liu2024LostMiddle,he2024retrieving}.

In contrast, \gls{kant} integrates \emph{knowledge anchors}, internal associations between question and code context, directly into the \gls{llm}'s parameters. 
This design reduces token usage at inference, improves precision, and avoids external dependency. 
Similar to \citeauthor{alassan2024benchmarking}~\cite{alassan2024benchmarking}, who demonstrate the viability of open-source \glspl{llm} for on-premise deployment in privacy-sensitive settings, \gls{kant} supports fully local fine-tuning and inference to ensure data confidentiality and deployment autonomy.

\textit{Parameter-efficient fine-tuning.}
To minimize the computational cost of repository-specific adaptation, we fine-tune all models using QLoRA, a quantization-aware variant of Low-Rank Adaptation (LoRA)~\cite{hu2022lora} that enables efficient fine-tuning on 4-bit quantized weights with minimal memory overhead~\cite{dettmers2023QLoRa}.
QLoRA combines trainable low-rank updates with paged optimizers to preserve performance while operating under strict hardware constraints, making it well-suited for large models in limited-resource environments.
We use QLoRA for fine-tuning all models to maximize efficiency without compromising performance.

\section{Future work}\label{sec:futureWork}
This section discusses implications and several promising directions that remain for future research.

An improvement direction is refining the granularity of knowledge anchors beyond the current file level. Anchoring at finer levels, such as methods, class or type declarations, or control-flow blocks could improve retrieval and reasoning over semantically localized code regions. 
This is especially useful for answering developer questions tied to detailed implementation semantics, including best practices, edge-cases, or parameter-specific logic. Prior work has shown the prevalence of such questions and their reliance on small but critical code elements, including method preconditions, control flow paths, and usage constraints \cite{latoza2010hard,sillito2006Questions,fritz2010using}. Finer-grained anchors could thus improve both retrieval accuracy and response specificity by enabling more precise access to structurally rich scoped code elements.

Another promising direction enriching knowledge anchors with contextual metadata from auxiliary artifacts, such as version control, issue trackers, or ownership graphs. This would enable reasoning about authorship, design intent, and maintenance activity, aspects not captured in source code alone. Such questions have been observed in prior work~\cite{fritz2010using}, yet remain difficult to address in \gls{kant}'s current setup, as it relies purely on source code. Augmenting anchors with this metadata could help the model answer historically and organizationally grounded queries otherwise inaccessible.
A further avenue involves adapting \gls{kant} to reasoning-centric / chain-of-thought (CoT) models. 
Integrating knowledge anchors with CoT prompting or structured memory access could support deeper, multi-hop reasoning over complex repositories, improving coherence across answers and promoting more consistent retrieval of interconnected knowledge.

The generalizability of \gls{kant} across programming languages also warrants further investigation. 
While the current evaluation focused on a Python-based repository, extending the approach to languages such as JavaScript, C++, and Java will help assess whether knowledge anchors remain stable and precise across different syntactic and semantic structures. 
Language-specific benchmarks will be essential for evaluating \gls{kant}'s robustness in more structurally diverse and syntactically varied environments.

Supporting evolving codebases is another key concern. 
To maintain model accuracy over time, we plan to explore incremental fine-tuning and memory update strategies that allow \gls{kant} to adapt to repository changes without requiring full retraining, allowing CI/CD integration.
Lastly, a deployment toolkit is planned to facilitate ease of adoption.

Collectively, these directions will further improve the utility of \gls{kant}, enabling more intelligent, scalable, and context-aware developer tools grounded in \glspl{llm}.

\section{Threats to Validity}\label{sec:threats}
We outline the potential threats to the validity of our findings and describe the steps taken to mitigate them.

\textit{Construct validity.}
Construct validity concerns whether the evaluation metrics and procedures accurately capture the intended phenomena, in this case, \gls{qa} quality and comprehension efficacy. 
Since human evaluations inherently involve subjective interpretation, we mitigated this risk by providing detailed scoring rubrics and assigning each answer to one of three independent human evaluators. 
A LocalStack domain expert independently rated the same set of responses without access to reviewer scores, providing an additional calibration point. 
For \rqa~ and \rqb, \gls{qa} pairs were randomly assigned to evaluators to minimize bias introduced by evaluator familiarity or systematic ordering effects.

\textit{Internal validity.}
The quality of generated \gls{qa} data and the downstream performance of fine-tuned models are influenced by the design of the prompts used in generation. 
To minimize this threat, we conducted a preliminary prompt optimization process through trial and error. 
For instance, we empirically found that splitting the question generation task into two targeted subprompts, one for general understanding and one for search related comprehension tasks, improved specificity and coherence. 
Although this iterative refinement was not formalized as a chain-of-thought search, it followed a progressive, outcome-guided tuning process. 
Nevertheless, some residual bias in prompt formulation may persist.

\textit{External validity.}
Our experiments focus exclusively on the LocalStack codebase, a large-scale, open-source Python project representative of modern cloud infrastructure systems. 
While this provides a realistic and high-impact evaluation context, generalization to other ecosystems, such as statically typed languages (e.g., Java, C++) or front-end focused languages (e.g., JavaScript), remains untested. 
We leave cross-language replication and multi-domain benchmarking as important directions for future work.

\textit{Conclusion validity.}
Although we followed standard statistical practices in determining sample sizes (e.g., 95\% confidence with 5\% margin of error), 
practical constraints such as annotator availability limited the absolute scale of human evaluation. 
We addressed this by combining reviewer ratings with independent expert assessments, and by using a structured Likert scale framework to collect detailed judgments across dimensions. 
We deliberately excluded automated \gls{qa} metrics (e.g., BLEU, ROUGE) due to their known misalignment with human judgment in code-oriented tasks~\cite{shi2023sotana,hicke2023chata}.

\textit{Tooling limitations.}
The \gls{rag} baseline depends on both the embedding model and the retrieval backend. 
While AoE embeddings~\cite{li-li-2024-aoe} offer a strong, lightweight baseline, performance may vary with different embedding methods or different chunking strategies. 

\textit{Data generation bias.}
All \gls{qa} data was synthesized using \llama-2-Chat 7B. 
Although the model provides high-quality generation capabilities, it reflects limitations of its training data and decoding strategies. 
Employing more recent or specialized code-generation models could yield better synthetic data and potentially stronger downstream performance. 
Exploring this axis remains a compelling avenue for future work.

\textit{Model compatibility.}
The \gls{kant} approach relies on \gls{fim} capabilities, which limits compatibility to models such as \codellama~ and \codegranite. 
Adapting \gls{kant} to models without \gls{fim} support would require architectural or pretraining modifications~\cite{bavarian2022efficienttraininglanguagemodels}, 
which may impact ease of adoption in practice.

\section{Conclusion}\label{sec:conclusion}
This work introduces \gls{kant}, a novel, lightweight, and efficient approach for repository-specific code comprehension in \glspl{llm}. \gls{kant} advances automated knowledge sharing in software repositories through:
\begin{enumerate*}[label=(\roman*)]
\item a validated and scalable pipeline that synthesizes instruction-tuning datasets directly from source code. These datasets, composed of \gls{qa} pairs, eliminate reliance on outdated, misaligned, or labor-intensive manually constructed datasets, reducing the need for extensive human supervision, and

    \item a fine-tuning strategy that embeds symbolic knowledge anchors into both training and 
    inference, enabling models to retrieve semantically grounded, repository-internal context without dependence on external documentation, long in-context prompts, or excessive token overhead.
\end{enumerate*}

Evaluations with \codellama~ and \codegranite, showed the effectiveness of this approach.
Most notably, the LocalStack expert preferred \gls{kant}-generated answers in over 79\% of cases, highlighting strong repository specific alignment. 
Human annotators likewise favored \gls{kant} in over 60\% of comparisons, confirming its effectiveness in repository level code comprehension.
These preferences were further supported by Likert scale evaluations from both the expert and annotators, indicating consistent gains in completeness, usefulness, and accuracy.
Compared to the \gls{sota} baseline, \gls{kant} consistently produced more coherent and semantically grounded responses.
In addition, \gls{kant} reduced inference latency by up to 85\%. These findings indicate a suitability for low latency, on premise environments.

\gls{kant} guides the model to develop localized memory regions that align with semantically meaningful code units, thereby increasing the likelihood that subsequent token predictions remain anchored to relevant knowledge. This enables repository-specific comprehension without the token overhead associated with long in-prompt contexts. 

By mitigating attention saturation and reducing susceptibility to hallucinations, this strategy provides a lightweight yet effective mechanism for enabling precise and efficient inference in \glspl{llm}.

In summary, \gls{kant} presents a practical and efficient \gls{llm}-based pipeline for code comprehension.
By embedding knowledge anchors directly into the model, it improves answer informativeness and semantic precision while reducing inference costs.
These contributions establish \gls{kant} as a practical and scalable foundation for precise and context-aware developer tools.

\section*{Acknowledgments}

The research leading to these results has been supported by the Swiss National Science Foundation (SNSF) through Grant SNSF204632.

\bibliographystyle{elsarticle-num-names} 
\bibliography{reference.bib}
\end{document}